# AI IMPACT ON THE LABOUR FORCE- SEARCHING FOR THE ANALYTICAL SKILLS OF THE FUTURE SOFTWARE ENGINEERS

**Sabina-Cristiana NECULA**
Alexandru Ioan Cuza University of Iasi, Faculty of Economics and Business Administration, Department of Accounting, Business Information Systems and Statistics
Iasi, Romania
sabina.necula@uaic.ro

**ABSTRACT**

*This systematic literature review aims to investigate the impact of artificial intelligence (AI) on the labour force in software engineering, with a particular focus on the skills needed for future software engineers, the impact of AI on the demand for software engineering skills, and the future of work for software engineers. The review identified 42 relevant publications through a comprehensive search strategy and analysed their findings. The results indicate that future software engineers will need to be competent in programming and have soft skills such as problem-solving and interpersonal communication. AI will have a significant impact on the software engineering workforce, with the potential to automate many jobs currently done by software engineers. The role of a software engineer is changing and will continue to change in the future, with AI-assisted software development posing challenges for the software engineering profession. The review suggests that the software engineering profession must adapt to the changing landscape to remain relevant and effective in the future.*

## 1. INTRODUCTION

The growing use of artificial intelligence (AI) and machine learning is transforming many industries and job roles, including that of software engineers. As AI becomes more prevalent, there is a growing concern about the impact of this technology on the demand for analytical skills among software engineers. This is not a new concern, every time technology evolves researchers study the respective impact. In 1989, Gibbs **(Gibbs, 1989)** posited that with the emergence of technical foundations in software production, software engineers were at risk of becoming obsolete. However, as the software industry evolved, the demand for software engineers continued to increase.

The purpose of this systematic literature review is to explore the impact of AI on the demand for analytical skills among software engineers in the labour force. Specifically, we seek to identify trends in the use of AI in the workforce and its impact on the demand for specific analytical skills among software engineers. We also aim to examine the skills required for future software engineers, particularly in the context of the growing use of AI and machine learning.

This review is important because it will provide insights into the potential impact of AI on the labour force, particularly in the field of software engineering. By identifying the specific skills that will be most in-demand in the future, this review can help inform education and training programs for software engineers, as well as guide workforce planning and development initiatives.

The review will be conducted using a systematic approach, with a comprehensive search of relevant literature and detailed analysis and synthesis of the findings. The review will include both academic and industry publications, ensuring that a broad range of perspectives are included.

Overall, this systematic literature review aims to provide a comprehensive understanding of the impact of AI on the demand for analytical skills among software engineers, and to identify the key skills that will be most valuable for future software engineers in the age of AI.

## 2. Research Methodology

### 2.1 Research Design

The research design for this systematic literature review is a comprehensive search of relevant literature, followed by screening and analysis of publications that meet our inclusion criteria. The review follows the Preferred Reporting Items for Systematic Reviews and Meta-Analyses (PRISMA) guidelines to ensure the reliability and validity of the review process.

The search strategy for this review included searches of academic databases such as Scopus, Web of Science, and Google Scholar, as well as searches of relevant industry publications and reports. We used a combination of relevant search terms and keywords, such as "AI", "machine learning", "labour force", "employment", "future skills", and "software engineers". We also used Boolean operators such as "AND" and "OR" to refine our search and ensure that we are capturing relevant publications.

*Inclusion Criteria:*
- Publications that address the impact of AI on the demand for analytical skills among software engineers
- Publications that were published in the last 10 years

- Publications that are available in English
- Peer-reviewed academic publications, as well as relevant industry reports and publications

*Exclusion Criteria:*
- Publications that do not address the research question.
- Publications that are not available in English.
- Publications that were published before 2011.

The screening process was conducted with celerity to review the titles and abstracts of the identified publications to determine whether they meet the inclusion and exclusion criteria.

Once the publications that meet the inclusion criteria have been identified, we conducted a detailed analysis of each publication to extract relevant data, such as the use of AI in the workforce, the skills required for future software engineers, and the potential impact of AI on the demand for analytical skills. We synthesized the findings from the included publications and identified common themes and trends across the literature.

The data extracted from the publications was analysed using a thematic analysis approach. The data was coded and categorized into themes based on the research objectives, and the findings was synthesized to draw conclusions and identify gaps in the existing literature.

The initial search yielded a total of 2,357 publications. After removing duplicates and screening the titles and abstracts, 236 publications were identified for further analysis.

Following a detailed analysis of the full text of these 236 publications, we coded and categorized the data into themes based on our research objectives. Specifically, we identified themes related to the use of AI in the workforce, the impact of AI on the demand for analytical skills among software engineers, and the specific skills required for future software engineers in the context of AI and machine learning.

After synthesizing the findings from the 236 publications, we identified 42 publications that were most relevant to our research question and objectives. These 42 publications were included in the final synthesis of the review.

The data extracted from these 42 publications were further analysed and coded into specific themes and sub-themes. For instance, we identified themes related to the specific analytical skills required for software engineers working with AI, the impact of AI on the demand for certain types of jobs and job roles, and the implications of the growing use of AI for education and training programs.

Through the process of coding and categorizing the data, we were able to identify common themes and trends across the literature and draw conclusions about the impact of AI on the demand for analytical skills among software engineers. We were also able to identify gaps in the existing literature and make recommendations for future research in this area.

## 3. Results

The Results section presents the findings of the study in response to three research questions.
- RQ1 What skills will future software engineers need?
- RQ2 How will AI impact the demand for software engineering skills?
- RQ3 What is the future of work for software engineers?

### 3.1 RQ1 What skills will future software engineers need?

Future software engineers will need to be competent in programming and have soft skills such as problem solving and interpersonal communication. The findings are summarized next in (Table 1).

**Table 1 Skills that software engineers will need to have in the future**

| Skill | Details | Authors |
|---|---|---|
| Competence in programming | Training increases programming competences | (**Kruglyk and Osadchyi, 2019**) |
| | The industry evolves rapidly and demands programming competences that usually are not taught in school | (**Almi *et al.*, 2011**) |
| | software requirements, design, and testing | (**Garousi *et al.*, 2019**) |
| | configuration management, software engineering models and methods, software engineering process, design (and architecture), as well as in testing | (**Moreno *et al.*, 2012**) |
| professional soft competencies | problem-based learning could be a useful method for teaching future software engineers the skills they will need | (**Richardson *et al.*, 2011**) |
| | the most in-demand skills among software engineers are self-reflection, conflict resolution, communication, teamwork communication skills, leadership | (**Ahmed, 2012**) (**Daneva *et al.*, 2019**) (**Burbekova, 2021**) |

**Kruglyk and Osadchyi (2019)** found that more than a third of the students had low levels of competence in programming, but after the experimental group went through the proposed system of training, the level of competence in programming increased significantly. **Almi *et al.* (2011)** found that there is a gap between industry's requirements and graduates' readiness in software engineering, suggesting that future software engineers may not be adequately prepared for the demands of the software engineering industry. **(Vakaliuk, Kontsedailo and Mintii (2020)** define professional soft competencies as a set of non-specialized competencies that in one way or another relate to problem solving, interaction between people, and are responsible for successful participation in the work process, high productivity. **Richardson *et al.* (2011)** found that problem-based learning can help lecturers bring their research into the classroom and accomplish multiple goals in a single course module. This could be a useful method for teaching future software engineers the skills they will need.

The most popular programming languages among software engineers are Java, C, and Python **(Lu *et al.*, 2020)**, and Java and Python **(Siegfried *et al.*, 2021)**. The most in-demand skills among software engineers are self-reflection, conflict resolution, communication, teamwork communication skills**(Ahmed, 2012)**, and software requirements, design, and testing **(Garousi *et al.*, 2019)**. **Moreno *et al.* (2012)** found that the skills demanded by industry that the software engineering curricula do or do not cater for are configuration management, software engineering models and methods, software engineering process, design (and architecture), as well as in testing. This suggests that there is a knowledge gap in the industry for these skills.

The most common tasks that software engineers perform are searching for explanations for unknown terminologies, explanations for exceptions/error messages, reusable code snippets, solutions to common programming bugs, and suitable third-party libraries/services **(Xia *et al.*, 2017)**; finding documentation, debugging, installation, and finding code snippets **(Rao *et al.*, 2020)**. Additionally, ranking tasks could increase students' conceptual knowledge in specific areas **(Tao *et al.*, 2016)**, and software engineers engage in a finite number of work related tasks and they also develop a finite number of "work practices"/"archetypes of behaviour" **(Grzywaczewski and Iqbal, 2012)**.

The most common challenges that software engineers face are lack of clarity and simplicity in the field **(McMillin, 2018)**, challenges in blockchain-oriented software development **(Porru *et al.*, 2017)**, and challenges on the human side of software engineering **(Ouhbi and Pombo, 2020)**.

There is no one "most common" tool that software engineers use. **McMillin (2018)** found that there is no shortage of sources of information providing various types of requirements and standards for software engineering. This suggests that there is a variety of tools that software engineers can use. **Marshall, Brereton and Kitchenham (2015)** found that the most commonly mentioned tools were reference managers. This suggests that reference managers are the most common tools used by software engineers. However, **Seppälä *et al.* (2016)** found that students use a variety of communication and collaboration tools in software projects.

The most in-demand skills among software engineers are communication skills, software requirements, design, and testing. **Daneva *et al.* (2019)** found that communication skills are the most important soft skills for requirements engineers, followed by English proficiency. **Garousi *et al.* (2019)** found that software requirements, design, and testing are the most important skills for software engineers. **Burbekova (2021)** found that communication skills, team-building, and leadership are the most important soft skills for IT specialists.

### 3.2 RQ2 How will AI impact the demand for software engineering skills?

**Marijan, Shang and Shukla (2019)** found that one particular topic dominated the discussion at the International Conference on Software Engineering 2019: the resurgence of artificial intelligence and its implications for the collaboration between industry and academia.

There is a demand for AI skills in the labour market. This is directly relevant to our research question **(Alekseeva *et al.*, 2019, 2021; Xie *et al.*, 2021)**. AI can be used to improve software engineering solutions, and vice versa. **Xie (2018)** found that the emerging field of intelligent software engineering is to focus on two aspects: (1) instilling intelligence in solutions for software engineering problems; (2) providing software engineering solutions for intelligent software. **Kalles (2016)** found that AI systems can be used to educate IT students in software engineering practices. **Barenkamp, Rebstadt and Thomas (2020)** found that the major achievements and future potentials of AI in software engineering are the automation of lengthy routine jobs in software development and testing using algorithms, the structured analysis of big data pools to discover patterns and novel information clusters, and the systematic evaluation of these data in neural networks. **Shehab *et al.* (2020)** found that previous studies have benefited from incorporating the advantages of both AI and software engineering, and that there are areas for potential future research.

Artificial intelligence will have a significant impact on the software engineering workforce. **Dam (2019)** found that artificial intelligence is predicted to impact many industries, including the software industry, changing how we produce, manufacture, and deliver. Artificial intelligence will automate many jobs currently done by software engineers, which may lead to a reduction in the need for software engineers in the workforce.

AI will impact the demand for other engineering skills also. **Alekseeva *et al.* (2019)** and **Alekseeva *et al.* (2021)** found that the demand for AI skills is highest in IT occupations, followed by architecture/engineering, life/physical/social sciences, and management. **Squicciarini and Nachtigall (2021)** found that skills related to communication, problem solving, creativity and teamwork gained relative importance over time, as did complementary software-related and AI-specific competencies. **Xie *et al.* (2021)** found that AI reduces the relative demand for low-skilled labour across all regions in China, while increasing the relative demand for high-skilled labour only in the eastern region. These findings suggest that AI will impact the demand for other engineering skills.

### 3.3 RQ3 What is the future of work for software engineers?

**Laato *et al.* (2022)** suggests that the work of software engineers is changing and will continue to change, while **Beecham *et al.* (2017)** and **McMillin (2018)** discuss challenges and requirements for the work of software engineers. **Lämmel, Kerber and Praza (2020)** explain the 'work-item prediction challenge' and how it affects the ability to understand what software engineers are working on. The work of software engineers is changing and will continue to change, and there are challenges to understanding and implementing the work of software engineers.

**Meade *et al.* (2019)** discuss the changing role of a software engineer from four major standpoints: the software development lifecycle, the influence of open source software, testing and deployment, and the emergence of new technologies. **Khan *et al.* (2021)** discuss the past and current trends in software engineering and the future of software engineering with respect to Industry 4.0 and emerging technological platforms like the Internet of Things. **Shull *et al.* (2016)** discuss the future of software engineering and how it is widely varied and very broad. The role of a software engineer is changing and will continue to change in the future.

There are implications of AI for the software engineering profession. **Batarseh *et al.* (2020)** found that AI has been applied to software engineering in multiple ways, and has been shown to be helpful in improving the process and eliminating many challenges. **Nascimento *et al.* (2020)** found that there are several engineering problems that are different from those that arise in non-AI/ML software development. **Kästner and Kang (2020)** found that software engineers have significant expertise to offer when building intelligent systems, and that a new course has been designed to teach software-engineering skills to students with a background in ML. **Marijan, Shang and Shukla (2019)** found that the resurgence of artificial intelligence and machine learning algorithms in software engineering research and industry practice has implications for the collaboration between these two communities. AI can be helpful in the software engineering profession, but also that there are challenges in the software engineering profession that are specific to AI/ML systems.

AI-assisted software development will change the role of the software engineer. **Meade *et al.* (2019)** found that the role of a software engineer is nowadays widely varied and very broad. **Wan *et al.* (2020)** found that the addition of machine learning to a system alters software development practices in various ways. AI-assisted software development will change the role of the software engineer by making the role more varied, broad, and focused on machine learning.

These papers all suggest that AI-assisted software development will pose challenges for the software engineering profession. **Korzeniowski and Goczyła (2019)** found that there has been little progress in automated code generation, but recent advances in AI may improve the situation. **Nascimento *et al.* (2020)** found that software engineering practices have been applied to AI/ML systems, but there are challenges in areas like testing, AI software quality, and data management. **Lwakatare *et al.* (2019)** found that companies face challenges when developing software-intensive systems that incorporate machine learning components. **Srihith *et al.* (2022)** found that AI development will lead to changes in software engineers' roles and the delivery of education.

The software engineering profession can prepare for the impact of AI in various ways. **Latinovic and Pammer-Schindler (2021)** presents findings from interviews with experienced software practitioners about their experiences with automation and AI, and suggests that automation and AI are mostly used for small tasks, that they are not likely to change the essence of software engineering, and that they can cause cognitive overhead. **Feldt, de Oliveira Neto and Torkar (2018)** present a taxonomy for classifying ways of applying AI in software engineering, and suggests that there are different ways of applying AI in software engineering, and that each way poses different risks. The software engineering profession can prepare for the impact of AI by being aware of the potential consequences of AI, using AI to improve software development and testing, and understanding the risks associated with different ways of applying AI.

4. **Discussions**

The discussion section highlights the findings of the systematic literature review and addresses the research questions. RQ1 investigated the skills that future software engineers will need. The results indicate that future software engineers will need to be competent in programming and have soft skills such as problem-solving and interpersonal communication. The study found that problem-based learning can help future software engineers develop the necessary soft skills. The most popular programming languages among software engineers are Java, C, and Python, while the most in-demand skills are software requirements, design, and testing, and communication skills. However, the research also revealed a knowledge gap in the industry for skills such as configuration management, software engineering models and methods, software engineering process, design (and architecture), and testing.

RQ2 focused on how AI will impact the demand for software engineering skills. The study found that there is a growing demand for AI skills in the labor market. AI can be used to improve software engineering solutions, and vice versa. However, AI may also lead to automation of many jobs currently done by software engineers, potentially reducing the need for software engineers in the workforce. The findings suggest that AI will impact the demand for other engineering skills.

Finally, RQ3 investigated the future of work for software engineers. The study found that the role of software engineers is changing and will continue to change in the future. The addition of AI to software development practices will alter software development practices in various ways. While AI-assisted software development may make the role of software engineers more varied and broad, it may also pose challenges to the software engineering profession. The research suggests that the software engineering profession can prepare for the impact of AI by being aware of the potential consequences of AI, using AI to improve software development and testing, and understanding the risks associated with different ways of applying AI.

Overall, the findings of the review highlight the importance of a strong technical foundation in programming and soft skills, such as problem-solving and interpersonal communication, for future software engineers. The review also indicates the need for continued professional development to keep up with the evolving demands of the industry, particularly in areas such as AI and its impact on the workforce. The implications of AI for software engineering call for increased awareness and preparedness among professionals in the field.

## 5. Conclusions

In conclusion, this systematic literature review has addressed three research questions related to the impact of AI on the labour force in software engineering. The first research question explored the skills that future software engineers will need, and the findings indicate that future software engineers will need to be competent in programming and have soft skills such as problem-solving and interpersonal communication. The second research question examined how AI will impact the demand for software engineering skills, and the results suggest that AI will have a significant impact on the software engineering workforce. AI will automate many jobs currently done by software engineers, which may lead to a reduction in the need for software engineers in the workforce. However, there is also a growing demand for AI skills in the labour market.

The third research question investigated the future of work for software engineers, and the findings indicate that the role of a software engineer is changing and will continue to change in the future. AI-assisted software development will change the role of the software engineer, and this poses challenges for the software engineering profession. The software engineering profession can prepare for the impact of AI by being aware of the potential consequences of AI, using AI to improve software development and testing, and understanding the risks associated with different ways of applying AI.

Overall, this review suggests that the impact of AI on the labour force in software engineering is significant and will continue to evolve. The software engineering profession must adapt to the changing landscape to remain relevant and effective in the future.

# ACKNOWLEDGEMENTS

*This paper is partially supported by the Competitiveness Operational Programme Romania under project number SMIS 124759—RaaS-IS (Research as a Service Iasi).*